\documentclass[pre,amssymb,amsmath,preprintnumbers,nofootinbib,showpacs,12pt]{revtex4}

\usepackage{graphicx}
\usepackage{subfigure}
\usepackage{amsmath}
\usepackage{color}
\usepackage{verbatim}

\begin{document}

\title{Validity of nonequilibrium work relations for the rapidly expanding quantum piston}

\author{H. T. Quan}

\affiliation{Department of Chemistry and Biochemistry, \\
University of Maryland, College Park, MD 20742 USA}

\author{Christopher Jarzynski}

\affiliation{Department of Chemistry and Biochemistry, and Institute for Physical Science and Technology,\\
University of Maryland, College Park, MD 20742 USA}

\date{\today}

\begin{abstract}
Recent work by Teifel and Mahler [{\it Eur. Phys. J. B} {\bf 75}, 275 (2010)] raises legitimate concerns regarding the validity of quantum nonequilibrium work relations in processes involving moving hard walls.
We study this issue in the context of the rapidly expanding one-dimensional quantum piston.
Utilizing exact solutions of the time-dependent Schr\" odinger equation, we find that the evolution of the wave function can be decomposed into {\it static} and {\it dynamic} components, which have simple semiclassical interpretations in terms of particle-piston collisions.
We show that nonequilibrium work relations remains valid at any finite piston speed, provided both components are included, and we study explicitly the work distribution for this model system.
\end{abstract}


\maketitle


In the past two decades, much attention has been devoted to theoretical predictions and experimental investigations regarding the fluctuations of small systems away from thermal equilibrium.
These predictions include the nonequilbrium work relation~\cite{Jarzynski1997a,Jarzynski1997b}
\begin{equation}
\label{eq:nwr}
\langle e^{-\beta W} \rangle = e^{-\beta\Delta F},
\end{equation}
and the corresponding fluctuation theorem derived by Crooks~\cite{Crooks1998,Crooks1999,Crooks2000}
\begin{equation}
\label{eq:cft}
\frac{\rho_F(+W)}{\rho_R(-W)} = e^{\beta(W-\Delta F)},
\end{equation}
which pertain to the work ($W$) performed on a system driven out of equilibrium.
(See Ref.~\cite{Jarzynski2011} for details and a recent review of these and related results.)
Most of the research in this area has concerned systems evolving under classical deterministic or stochastic dynamics.
However, the past few years have seen increased focus on the fluctuations of quantum systems driven away from equilibrium~\cite{Campisi2011}.

\begin{figure}[tbh]
\includegraphics[bb=116bp 252bp 512bp 665bp, width=6cm]{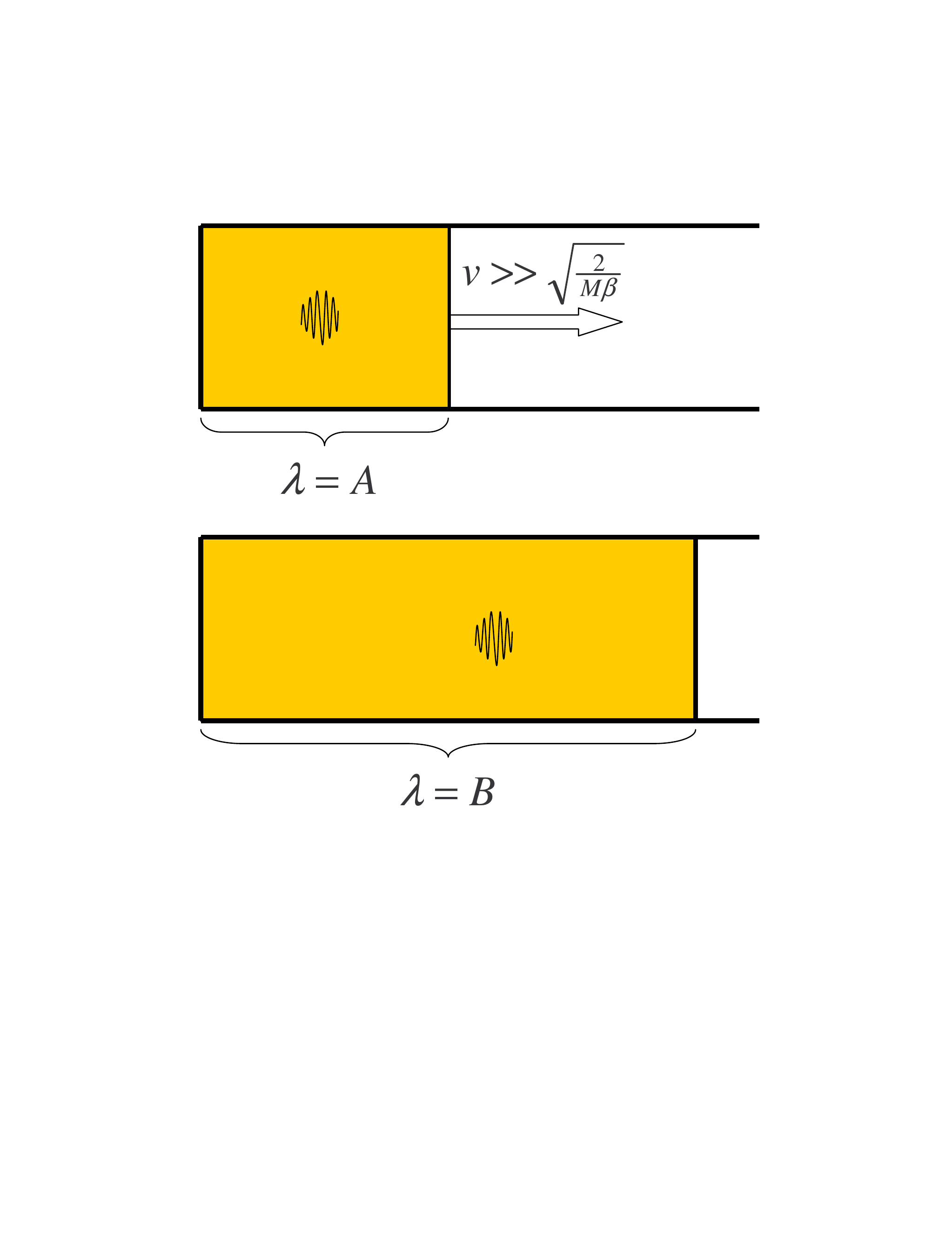}
\caption{Schematic depiction of a quantum piston. 
A quantum particle is confined by hard walls, one of which acts as an externally controlled piston.
We focus on the case in which the piston is pulled outward at a speed that is much greater than the initial thermal speed of the particle.
}
\label{piston}
\end{figure}

While the derivation of Eq.~\ref{eq:nwr} for an isolated quantum system is straightforward and rests on familiar properties of unitary evolution~\cite{Kurchan2000,Tasaki2000,Mukamel2003,Talkner2007}, Teifel and Mahler (TM)~\cite{Teifel2010} have recently presented a calculation suggesting that Eq.~\ref{eq:nwr} (and by extension, Eq.~\ref{eq:cft}) might be violated for the one-dimensional {\it quantum piston}.
In this familiar model system, the wavefunction describing a particle inside a box evolves in time as the length of the box is increased (Fig. \ref{piston}) or decreased.
Although TM focus specifically on this simple model, their analysis has broader implications, raising the possibility that Eqs.~\ref{eq:nwr} and \ref{eq:cft} might generically be violated for processes involving the motion of hard walls.
In such situations the system's Hilbert space changes with time, and questions of unitarity must be handled with care.
This feature has a classical counterpart, emphasized by Sung~\cite{Sung2008}:
the phase space accessible to a classical particle confined by hard walls changes with time as those walls move.

In the classical setting, processes involving moving boundaries have proven to be instructive~\cite{Lua2005,Presse2006,Jarzynski2006,Crooks2007,Sung2008}, deepening our understanding of nonequilibrium work relations by highlighting apparent paradoxes and counterintuitive features.
In this paper we use exact solutions of the time-dependent Schr\" odinger equation~\cite{Doescher1969} to investigate the validity of Eq.~\ref{eq:nwr} for an expanding quantum piston.

In what follows, we first sketch the usual derivation of Eq.~\ref{eq:nwr} for an isolated quantum system (Eqs.~\ref{eq:spectrum} - \ref{eq:sn_definition}), as well as an apparent counter-argument which suggests that Eq.~\ref{eq:nwr} is violated for the quantum piston (Eqs.~\ref{eq:chop} - \ref{eq:questionableLimit}).
We then apply the exact results of Ref.~\cite{Doescher1969} to the case in which the piston moves outward at speed $v$.
We find that Eq.~\ref{eq:nwr} is valid for any finite pulling speed, which seems to contradict the analysis in Eqs.~\ref{eq:chop} - \ref{eq:questionableLimit}.
We then consider the limit $v\rightarrow\infty$, and we find that the apparent discrepancy has an appealing semiclassical interpretation that parallels the purely classical analyses of Refs.~\cite{Lua2005,Presse2006,Jarzynski2006,Sung2008}.

Consider a quantum system whose parameter-dependent Hamiltonian $\hat H^\lambda$ has a discrete energy spectrum:
\begin{equation}
\label{eq:spectrum}
\hat H^\lambda \, \vert m^\lambda \rangle = E_m^\lambda \, \vert m^\lambda \rangle,
\end{equation}
with $m = 0, 1, 2\cdots$.
We use superscripts to indicate the value of the externally controlled parameter, $\lambda$, which for the case of the quantum piston is the position of the piston itself, equivalently the length of the box.
Now imagine that this system is subjected to the following process.
(1) With the parameter fixed at $\lambda=A$, the system is equilibrated with a reservoir at temperature $\beta^{-1}$; the reservoir is then disconnected and the energy of the system is measured.
At this point the system is in a pure state $\vert m^A \rangle$, set by the outcome of the energy measurement.
(2) The system now evolves under Schr\" odinger's equation, from time $t=0$ to $t=\tau$, as the parameter is varied from $A$ to $B$ according to a schedule, or protocol, $\lambda_t$.
The energy is then measured once more, resulting in ``collapse'' into an eigenstate $\vert n^B \rangle$ of $\hat H^B$.
Following Refs.~\cite{Kurchan2000,Tasaki2000,Mukamel2003,Talkner2007} we identify the work performed on the system with the net change in its energy:
\begin{equation}
\label{eq:Wdef}
W \equiv E_n^B - E_m^A.
\end{equation}
By repeating this process, we generate an ensemble of realizations, each defined by an initial state $\vert m^A\rangle$ and a final state $\vert n^B\rangle$.
The initial states are distributed according to
\begin{equation}
\label{eq:Peq}
P_m^{{\rm eq},A}= \frac{1}{Z_A} e^{-\beta E_m^A},
\end{equation}
where $Z_A = \sum_m e^{-\beta E_m^A} = e^{-\beta F_A}$ is the partition function, 
and the final states according to the conditional distribution
\begin{equation}
\label{eq:Pnm}
P(n^B \vert m^A) = \Bigl\vert \langle n^B \vert \hat U \vert m^A \rangle \Bigr\vert ^2,
\end{equation}
where $\hat U$ is the time-evolution operator that describes evolution under Schr\" odinger's equation from $t=0$ to $\tau$.
Combining Eqs.~\ref{eq:Wdef} - \ref{eq:Pnm}, the left side of Eq.~\ref{eq:nwr} can now be evaluated:
\begin{equation}
\label{eq:standardDerivation}
\langle e^{-\beta W} \rangle = 
\sum_m P_m^{{\rm eq},A} \sum_n P(n^B \vert m^A) \, e^{-\beta W} =
\frac{1}{Z_A} \sum_n e^{-\beta E_n^B} \, s_n,
\end{equation}
where
\begin{equation}
\label{eq:sn_definition}
s_n \equiv \sum_m P(n^B\vert m^A)
= \sum_m \langle n^B \vert \hat U \vert m^A \rangle \, \langle m^A \vert \hat U^\dagger \vert n^B \rangle.
\end{equation}
At this point, one normally argues that the sum $\sum_m \vert m^A \rangle \, \langle m^A \vert$ is the identity operator, hence $s_n=1$ and the right side of Eq.~\ref{eq:standardDerivation} becomes $Z_B/Z_A = e^{-\beta\Delta F}$, completing the proof.

Teifel and Mahler~\cite{Teifel2010} correctly point out that this argument requires care if the eigenstates of $\hat H^A$ do not span the Hilbert space of $\hat H^B$. For a quantum piston whose length is increased from $\lambda_0=A$ to $\lambda_\tau=B$ at speed $v$,
the states $\vert m^A \rangle$ are restricted to the interval $0 < x < A$,
whereas the final Hilbert space supports states extending over the wider interval $0 < x < B$.
If $\psi(x) = \langle x\vert\psi\rangle$ is a wave function belonging to the Hilbert space of $\hat H^B$, then the operator $\sum_m \vert m^A \rangle \, \langle m^A \vert$ effectively ``chops off'' a portion of this wavefunction:
\begin{equation}
\label{eq:chop}
 \sum_m \langle x \vert m^A \rangle \langle m^A \vert \psi \rangle = \theta(A-x) \, \psi(x) \, ,
\end{equation}
where $\theta(\cdot)$ is the unit step function.
We conclude that $\sum_m \vert m^A \rangle \, \langle m^A \vert$ {\it is not the identity operator} when it acts in the Hilbert space spanned by eigenstates of $\hat H^B$.
Hence the derivation described in the previous paragraph does not automatically apply to the quantum piston, and this raises concerns regarding the validity of Eq.~\ref{eq:nwr} in that context.

As a limiting case, let us analyze the {\it infinitely fast} expansion of the piston, $v\rightarrow \infty$.
The sudden approximation~\cite{Messiah1966} suggests that the wave function then remains in its initial state,
\begin{equation}
\label{eq:suddenApproximation}
\lim_{v\rightarrow\infty} \hat U \vert m^A \rangle = \vert m^A \rangle.
\end{equation}
Combining Eqs.~\ref{eq:sn_definition}-\ref{eq:suddenApproximation} leads to
\begin{equation}
\label{eq:questionableLimit}
\lim_{v\rightarrow \infty} s_n \overset{?}{=} 
\sum_{m=1}^\infty \langle n^B \vert m^A \rangle \, \langle m^A \vert n^B \rangle
= \int_0^{A} {\rm d}x \, \left\vert \phi_n(x;B) \right\vert^2
= \frac{1}{r} - \frac{\sin(2\pi n /r)}{2\pi n} < 1,
\end{equation}
where $r \equiv B/A$ and the wavefunction
\begin{equation}
\label{eq:phi}
\phi_n(x;\lambda) = \sqrt{\frac{2}{\lambda}} \sin \left( \frac{n\pi x}{\lambda} \right)
\end{equation}
describes the $n$'th eigenstate of $\hat H^\lambda$.
(The notation $\overset{?}{=}$ indicates that we question the validity of the first step in Eq.~\ref{eq:questionableLimit}.)
Substitution of Eq.~\ref{eq:questionableLimit} ($s_n<1$) into Eq.~\ref{eq:standardDerivation} implies a violation of Eq.~\ref{eq:nwr}.
In the opposite limit, namely {\it adiabatic} expansion, $v\rightarrow 0$, TM find that Eq.~\ref{eq:nwr} is satisfied.
These considerations suggest that for the expansion of a quantum piston at finite speed $v$, Eq.~\ref{eq:nwr} is only approximately valid, but the approximation becomes exact in the adiabatic limit, $v\rightarrow 0$.

In what follows we will argue that in fact $s_n=1$ {\it for all finite values of} $n$ {\it and} $v$, and therefore
\begin{equation}
\lim_{v\rightarrow \infty} s_n = 1,
\end{equation}
in contradiction with Eq.~\ref{eq:questionableLimit}.
By Eq.~\ref{eq:standardDerivation}, our conclusion implies that Eq.~\ref{eq:nwr} is valid for any finite piston speed.

For a quantum piston expanding at speed $v$ from an initial length $\lambda_0=A$,
a set of independent solutions to the time-dependent Schr\"odinger equation can be written as~\cite{Doescher1969}
\begin{equation}
\label{solution}
\Phi_{l}(x,t)=
\exp\left[ \frac{i}{\hbar\lambda_t} \left(
\frac{1}{2} Mvx^2 - E_l^A At \right)
\right] \,
\phi_l(x;\lambda_t),
\quad\quad
l = 1, 2,  \cdots,
\end{equation}
where $M$ denotes the mass of the particle, and $E_l^A = l^2 \pi^2\hbar^2/2MA^2$ is the $l$'th eigenenergy of the system at $t=0$.
The wavefunctions $\Phi_{l}(x,t)$ form a complete
orthonormal set, $\langle \Phi_k \vert \Phi_l \rangle = \delta_{kl}$, but are not eigenstates of $\hat H^{\lambda_t}$.
(The $\phi_l$'s defined in Eq.~\ref{eq:phi} are the eigenstates.)
A general solution of the time-dependent Schr\"odinger equation takes the form
\begin{equation}
\Psi(x,t)=\sum_{l=1}^{\infty}c_{l} \, \Phi_{l}(x,t),
\end{equation}
where the {\it time-independent} coefficients $c_{l}$ are set by the initial wave function:
\begin{equation}
c_{l} =\int_{0}^{A} \Phi_{l}^{*}(x,0) \Psi(x,0) \, {\rm d}x.
\end{equation}
For initial conditions
$\vert \Psi(0) \rangle = \vert m^A \rangle$
these coefficients are (setting $\hbar = M = 1$)
\begin{subequations}
\label{eq:exactSoln}
\begin{equation}
\label{eq:clm}
c_{l}(m)=\frac{2}{A}\int_{0}^{A}e^{-i vx^2/2 A}
\,
\sin \left({\frac{l\pi x}{A}} \right) \sin \left({\frac{m\pi x}{A}}\right) \, {\rm d}x,
\end{equation}
and the transition matrix element to the state
$\vert n^B \rangle$
at the final time $\tau$ is
\begin{equation}
\label{transition}
\langle n^B \vert \hat U \vert m^A \rangle
= \left\langle n^B \vert \Psi(\tau) \right\rangle
=\sum_{l=1}^{\infty} c_{l}(m) \int_{0}^B
\phi^{*}_{n}(x;B) \, \Phi_{l}(x,\tau) \, {\rm d}x.
\end{equation}
\end{subequations}
Eqs.~\ref{eq:Pnm} and \ref{eq:exactSoln} give the transition probability $P(n^B\vert m^A)$, in terms of one-dimensional integrals that are easily computed numerically.
This transition probability satisfies normalization:
\begin{equation}
\label{eq:normF}
\sum_n P(n^B\vert m^A) =
\int_0^B {\rm d}x\, \left\vert \langle x \vert \hat U \vert m^A \rangle \right\vert^2 = 1.
\end{equation} 

Although we have considered the expansion of a quantum piston, Eq.~\ref{solution} is equally valid for compression~\cite{Doescher1969}.
By reversing the roles of $A$ and $B$ and the roles of $m$ and $n$ and by replacing $v$ with $-v$ (in Eq.~\ref{eq:exactSoln}) we obtain the transition probability $\bar{P}(m^A\vert n^B)$ from the $n$'th eigenstate of $\hat H^B$ to the $m$'th eigenstate of $\hat H^A$, where the notation $\bar{P}$ indicates the compression process.
This transition probability also satisfies normalization:
\begin{equation}
\label{eq:normR}
\sum_m \bar{P}(m^A\vert n^B) =
\int_0^A {\rm d}x\, \left\vert \langle x \vert \hat U^\prime \vert n^B \rangle \right\vert^2 = 1,
\end{equation}
where $\hat U^\prime$ is the time-evolution operator for the compression process.

In the Appendix, we provide explicit expressions for $P(n^B\vert m^A)$ and $\bar{P}(m^A\vert n^B)$, and using these expressions we directly verify the relation
\begin{equation}
\label{symmetry}
P(n^B\vert m^A)=\bar{P}(m^A\vert n^B).
\end{equation}
It should be clear that this relation is precisely what we expect from time-reversal invariance ($\hat U^\prime = \hat U^\dagger$), see e.g.\ Eq.~56 of Ref.~\cite{Campisi2011}.
Using Eq.~\ref{symmetry} we can now transform the sum over initial states in Eq.~\ref{eq:sn_definition} into a sum over final states:
\begin{equation}
\label{eq:sum_to_unity}
s_n \equiv \sum_m P(n^B \vert m^A) = \sum_m \bar{P}(m^A\vert n^B) = 1
\end{equation}
using Eq.~\ref{eq:normR} in the last step.
Since this result is independent of $v$, we conclude that Eq.~\ref{eq:nwr} is valid at any finite speed of expansion.

To obtain Eq.~\ref{eq:cft} by similar means, we follow Tasaki~\cite{Tasaki2000} and write explicit expressions for the forward and reverse work distributions (corresponding to piston expansion and compression, respectively):
\begin{equation}
\label{eq:derivecft}
\begin{split}
\rho_F(W) &= Z_A^{-1} \sum_m e^{-\beta E_m^A} \sum_n P(n^B \vert m^A) \, \delta \left( W - E_n^B + E_m^A \right), \\
\rho_R(W) &= Z_B^{-1} \sum_n e^{-\beta E_n^B} \sum_m \bar{P}(m^A \vert n^B) \, \delta \left( W - E_m^A + E_n^B \right).
\end{split}
\end{equation}
For every realization $m^A \rightarrow n^B$ that gives a particular work value during the forward process, there is a corresponding realization $n^B \rightarrow m^A$ that gives the opposite work value during the reverse process.
Combining this observation with Eqs.~\ref{symmetry} and \ref{eq:derivecft} we obtain Eq.~\ref{eq:cft}~\cite{Tasaki2000}.

\begin{figure}[tbp]
   \begin{center}
      \subfigure[$\,\,v=10$]{
        \label{fig:v=10}
         \includegraphics[trim = 0in 0in 0in 0in , scale=0.29,angle=270]{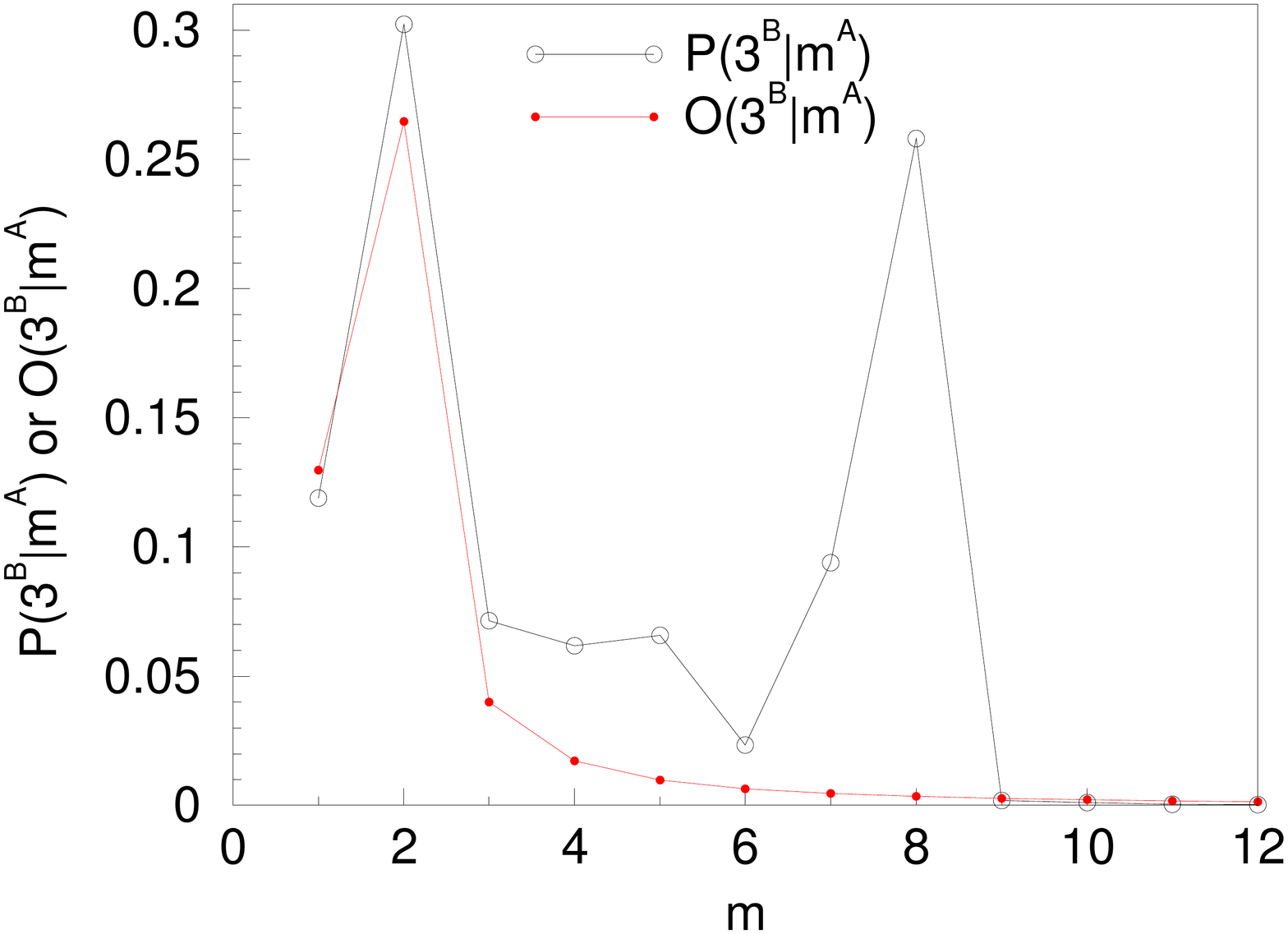} 
      }
      \subfigure[$\,\,v=100$]{
         \label{fig:v=100}
         \includegraphics[trim = 0in 1in 0in 0in , scale=0.29,angle=270]{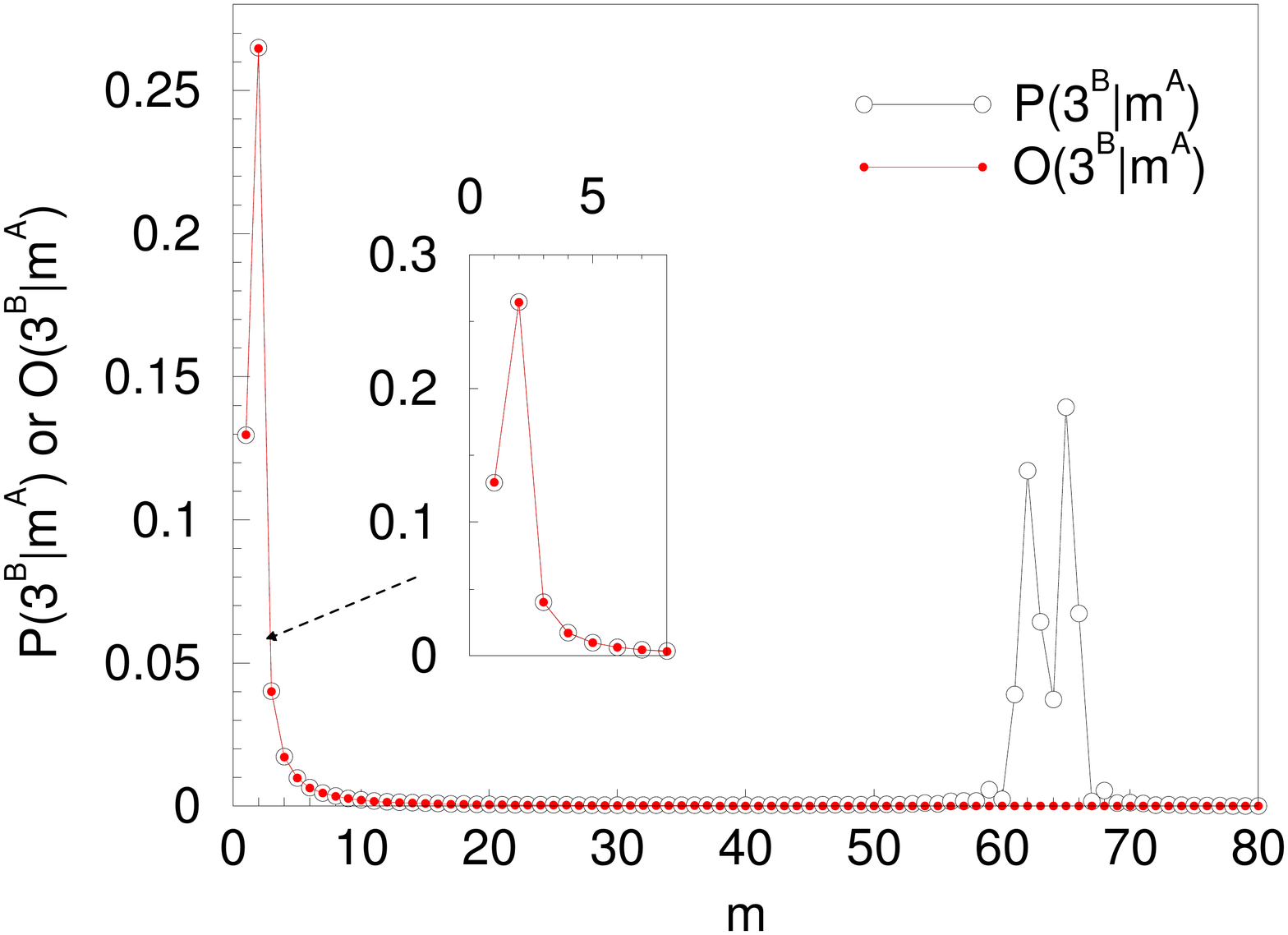} 
      } 
      \subfigure[$\,\,v=500$]{
         \label{fig:v=500}
         \includegraphics[trim = 0in 1in 0in 0in , scale=0.29,angle=270]{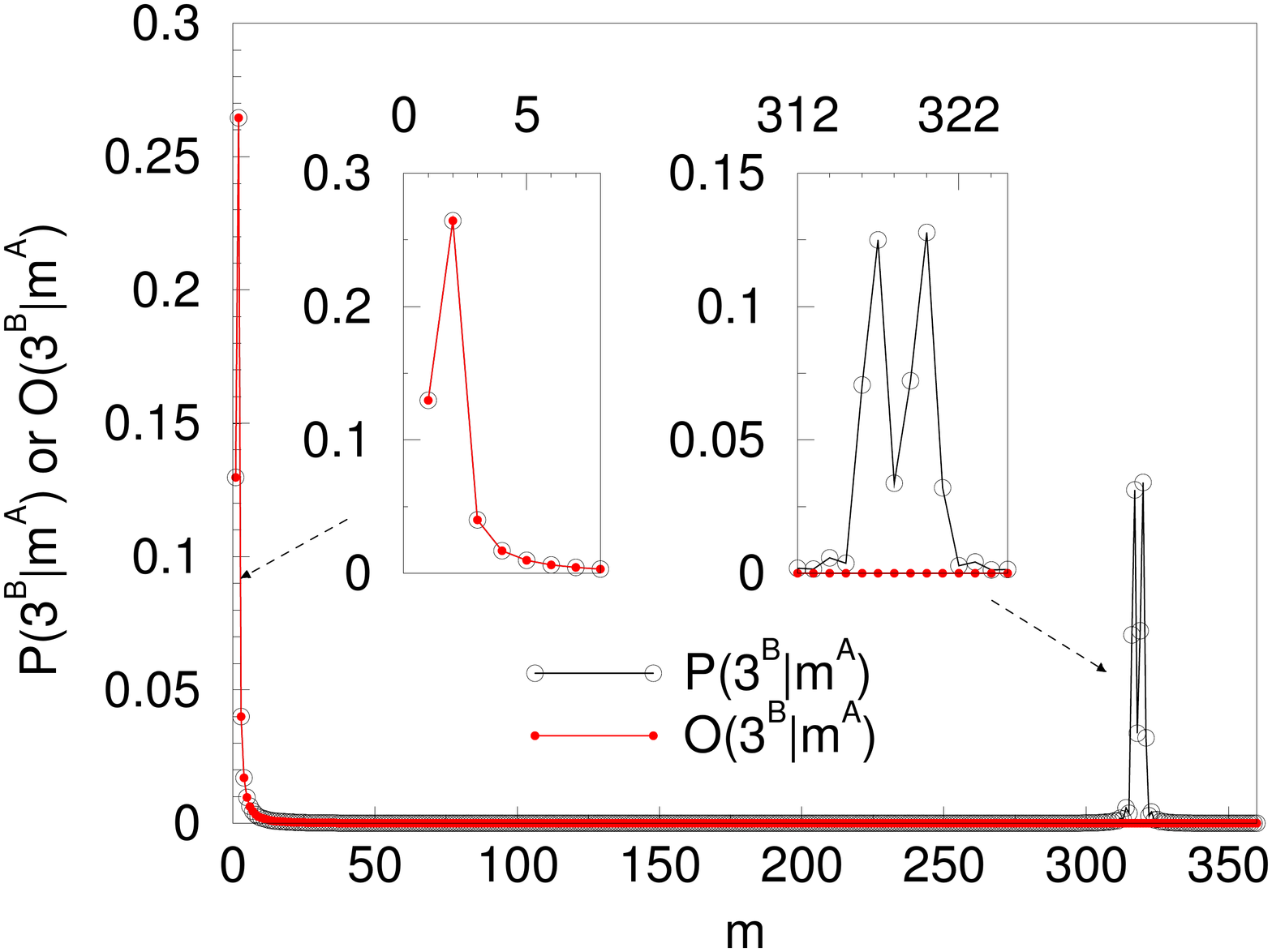} 
      } 
   \end{center}
   \caption{$P(n^B \vert m^A)$ 
   is plotted as a function of $m$ at fixed $n=3$ (open circles), revealing a two-peak structure, with the left peak around $m = 2$ and the right peak near $m = 2vA/\pi$.
   We refer to these peaks as the static and dynamic components, respectively.
   Also plotted is the quantity $O(n^B \vert m^A)$ (red points), which displays only a single peak around $m = 2$.
   Note that for $v=100$ and $v=500$ the single peak of $O(n^B \vert m^A)$ is virtually identical to the static component of $P(n^B \vert m^A)$.
   }
    \label{fig:peaks}
\end{figure}

Up to this point we have used the symmetry relation, Eq.~\ref{symmetry}, to show that $s_n=1$ for any finite speed $v$, and therefore that Eqs.~\ref{eq:nwr} and \ref{eq:cft} remain valid for the quantum piston.
However, this analysis does not yet explain why Eq.~\ref{eq:questionableLimit} gives a contradictory result in the limit $v\rightarrow\infty$.
To address this issue, in the following paragraphs we present numerical evidence that the value of $s_n$ is naturally expressed as the sum of a static and a dynamic contribution, reflected in the two-peak structure seen in Figs.~\ref{fig:v=10} - \ref{fig:v=500}.
The sum of these contributions is unity for any finite $v$ (as per Eq.~\ref{eq:sum_to_unity}), but Eq.~\ref{eq:questionableLimit} {\it accounts only for the static contribution}, thus giving $s_n^{\rm Eq.(\ref{eq:questionableLimit})}<1$.
Here and in the following discussion, we use the notation $s_n^{\rm Eq.(\ref{eq:questionableLimit})}$ to denote the value for $s_n$ predicted (incorrectly!) by Eq.~\ref{eq:questionableLimit}, in the limit $v\rightarrow\infty$. 
After presenting the numerical results, we suggest a semiclassical interpretation in terms of piston-particle collisions.

We have used Eq.~\ref{eq:exactSoln} to evaluate $P(n^B \vert m^A)$ numerically.
In Fig.~\ref{fig:peaks}, this quantity is plotted for fixed final state $n=3$, as a function of initial state $m=1, 2, \cdots$, for piston expansion from $A=1.0$ to $B=2.0$ at various speeds: $v=10$, 100 and 500.
The plot reveals a two-peak structure.
The left peak, near $m=2$, remains approximately independent of $v$, whereas the right peak is located near $m = 2vA/\pi$; thus with increasing $v$ the right peak shifts further rightward.
(Note the change of scale in the plots.)
We will refer to the left and right peaks as the {\it static} and {\it dynamic} components, respectively.
We can decompose the value of $s_n$, with $n=3$ in our case, into contributions from these components:
\begin{equation}
s_n^L = \sum_{m\le m^*} P(n^B\vert m^A),
\quad\quad
s_n^R = \sum_{m > m^*} P(n^B\vert m^A).
\end{equation}
Here $m^*$ is the value of $m$ at which $P(n^B\vert m^A)$ is minimized in the region between the two peaks.
Table~\ref{table:sn} lists the values of these contributions, obtained by numerical evaluation of the integrals in Eq.~\ref{eq:exactSoln}, as well as their sum, $s_n$.
Note that $s_n=1.000$ at all three speeds, in agreement with Eq.~\ref{eq:sum_to_unity}.

\begin{table}[h!]
\begin{tabular}{c||c|c|c|c|}
~ &  $v=10$ & $v=100$ & $v=500$ & $v\rightarrow\infty$ \\
\hline
$s_n^{\rm Eq.(\ref{eq:questionableLimit})}$ &  &  &  & 0.500 \\
\hline
$s_n^L$ & ~0.644 & ~0.499 & ~0.500 & \\
$s_n^R$ & 0.356 & 0.501 & 0.500 & \\
\hline
$s_n = s_n^L + s_n^R$ & 1.000 & 1.000 & 1.000 &
\end{tabular}
\caption{Static ($L$) and dynamic ($R$) contributions to $s_{n=3}$, as well as the asymptotic value of $s_n$ predicted by Eq.~\ref{eq:questionableLimit}, for piston expansion from $A=1.0$ to $B=2.0$.}
\label{table:sn}
\end{table}

Let us now rewrite Eq.~\ref{eq:questionableLimit} as
\begin{equation}
\lim_{v\rightarrow\infty} s_n\overset{?}{=} \sum_m \langle n^B \vert m^A \rangle \, \langle m^A \vert n^B \rangle
= \sum_m \left\vert \int_0^A {\rm d}x \, \phi_n^*(x;B) \phi_m(x;A) \right\vert^2
\equiv \sum_m O(n^B\vert m^A).
\end{equation}
We can interpret the overlap $O(n^B\vert m^A) = \left\vert \langle n^B \vert m^A \rangle \right\vert^2$ as the probability to end in state $\vert n^B\rangle$ after the measurement of the final energy, when starting from state $\vert m^A\rangle$, {\it under the assumption that the wave function remains unchanged during the sudden expansion}.
This assumption amounts to a literal intepretation of the sudden approximation, Eq.~\ref{eq:suddenApproximation}.
Using Eq.~\ref{eq:phi} to evaluate the integral, in Fig.~\ref{fig:peaks} we have also plotted $O(n^B\vert m^A)$, which exhibits a single peak around $m=2$.
We observe that the larger the value of $v$, the more closely $O(n^B\vert m^A)$ resembles the left peak of $P(n^B \vert m^A)$; indeed at $v=100$ and 500 they are virtually identical.
These empirical observations suggest that Eq.~\ref{eq:questionableLimit} captures only the contribution to $s_n$ from the static component $s_n^L$, while missing the contribution from the dynamic component $s_n^R$.

Quantitatively, $s_n^{\rm Eq.(\ref{eq:questionableLimit})} = 0.5$ for $A=1.0$, $B=2.0$, and $n=3$, whereas the data in Table~\ref{table:sn} suggest that the static contribution $s_n^L$ approaches $0.5$ as $v\rightarrow\infty$.
Moreover, Table~\ref{table:sn_r=1.485} lists these quantities for the case $A=1.0$, $B=1.485$, and $n=3$, with $s_n^L$ and $s_n^R$ again calculated using Eq.~\ref{eq:exactSoln}.
Once again we find that $s_n^L + s_n^R = 1.000$ at all speeds, and $s_n^L \rightarrow s_n^{\rm Eq.(\ref{eq:questionableLimit})} \approx 0.667$ as $v\rightarrow\infty$.
These findings are consistent with our hypothesis that Eq.~\ref{eq:questionableLimit} reflects only the static and not the dynamic contribution to $s_n$.

\begin{table}[h!]
\begin{tabular}{c||c|c|c|c|}
~ &  $v=10$ & $v=100$ & $v=500$ & $v\rightarrow\infty$ \\
\hline
$s_n^{\rm Eq.(\ref{eq:questionableLimit})}$ &  &  &  & 0.667 \\
\hline
$s_n^L$ & ~0.638 & ~0.667 & ~0.667 & \\
$s_n^R$ & 0.362 & 0.333 & 0.333 & \\
\hline
$s_n = s_n^L + s_n^R$ & 1.000 & 1.000 & 1.000 &
\end{tabular}
\caption{Same as Table~\ref{table:sn}, but for expansion from $A=1.0$ to $B=1.485$.}
\label{table:sn_r=1.485}
\end{table}

We now build a semiclassical interpretation to reinforce these conclusions.
For $A=1.0$, $B=2.0$, and piston speed $v=100$, consider the value $P(3^B \vert 64^A)$, corresponding to the right peak in Fig.~\ref{fig:v=100}.
This gives the probability to end in state $\vert 3^B\rangle$, starting from state $\vert 64^A\rangle$, during the expansion process.
Semiclassically, the initial state $\vert \Phi(0) \rangle = \vert 64^A\rangle$ can be imagined as a particle moving with speed
\begin{equation}
\label{eq:u}
\vert u\vert = \sqrt{2E_{m=64}^A} = \frac{m\pi}{A} \approx 200
\end{equation}
between two hard walls.
At $t=0$, when the piston begins to move rightward with speed $v=100$, the particle is moving either leftward ($u \approx -200$) or rightward ($u \approx +200$), with equal likelihood.
In the latter case, the particle will collide once with the receding piston, losing approximately all of its kinetic energy.
The final state $\vert \Phi(\tau) \rangle$ will then contain a substantial component of low-energy states (including $\vert 3^B\rangle$) reflecting this one-collision scenario.
In other words, $P(3^B \vert 64^A)$ is non-negligible because a single collision with the piston scatters the particle from the high-energy state $\vert 64^A\rangle$ to the low-energy state $\vert 3^B\rangle$.
The same argument explains, quantitatively, why the right peak occurs at $m^A \approx 320$ in Fig.~\ref{fig:v=500}.

Alternatively, we can use Eq.~\ref{symmetry} to rewrite $P(3^B \vert 64^A)$ as $\bar{P}(64^A \vert 3^B)$, which is the probability to {\it end} in state $\vert 64^A\rangle$, {\it starting} from state $\vert 3^B\rangle$ when compressing at piston speed $v=100$.
Here we imagine a particle initially moving with speed
\begin{equation}
\label{eq:u=5}
\vert u\vert = \sqrt{2E_{n=3}^B} \approx  5.
\end{equation}
As the piston moves from $B=2.0$ to $A=1.0$, the particle might suffer a single collision with the piston, imparting a leftward velocity $\Delta u \approx -2v = -200$ to the particle.
Thus for the initial state $\vert \Phi(0) \rangle = \vert 3^B\rangle$ we expect the final state $\vert \Phi(\tau) \rangle$ to be a superposition of low-energy states (corresponding to no collisions) and high-energy states near $\vert 64^A\rangle$ (one collision).
This is indeed the spectrum seen in Fig.~\ref{fig:v=100}.
This interpretation suggests that $s_n^R$ is equal to the probability that the particle suffers a collision with the piston during the compression process, and $s_n^L$ is the probability it avoids a collision, when starting from state $\vert n^B\rangle$.
Semiclassically and in the limit $v\rightarrow\infty$, the probability to avoid a collision during compression is just the probability to find the particle in the region $0<x<A$ at time $t=0$ (when the piston is at location $B$), which leads to
\begin{equation}
\label{eq:sc}
\lim_{v\rightarrow\infty}
s_n^{L,{\rm sc}} = \frac{A}{B} = \frac{1}{r} .
\end{equation}
The superscript ``sc'' emphasizes that this is a semiclassical approximation.
Eq.~\ref{eq:sc} agrees with the term $1/r$ in the expression appearing in Eq.~\ref{eq:questionableLimit} (just before the inequality); the oscillatory term there, $\sin(2\pi n/r)/2\pi n$, is quantum-mechanical in origin.

In either case -- expansion or compression -- the dynamic component is associated semiclassically with a collision between the particle and the piston.
We conclude that {\it Eq.~\ref{eq:questionableLimit} underestimates $s_n$ because it neglects the contribution due to a particle-piston collision}.

These considerations relate to the ordering of limits.
Fig.~\ref{fig:peaks} suggests that
\begin{equation}
\lim_{v\rightarrow\infty} P(n^B\vert m^A) = O(n^B\vert m^A),
\end{equation}
for any {\it fixed} initial state $\vert m^A\rangle$.
Now, Eq.~\ref{eq:questionableLimit} implicitly contains a double limit, namely,
\begin{equation}
\label{eq:limit1}
\lim_{v\rightarrow \infty} s_n =
\lim_{v\rightarrow \infty} \lim_{K\rightarrow\infty} \sum_{m=1}^{K} P(n^B \vert m^A).
\end{equation}
If we take the limit $K\rightarrow\infty$ first (with $v$ fixed), then both the static and dynamic components $s_n^L$ and $s_n^R$ are included in the sum, and the right side of Eq.~\ref{eq:limit1} sums to unity (Eq.~\ref{eq:sum_to_unity}):
\begin{equation}
\lim_{v\rightarrow \infty} \lim_{K\rightarrow\infty} \sum_{m=1}^{K} P(n^B \vert m^A) = 1.
\end{equation}
However, if we reverse the ordering of limits and first take $v\rightarrow\infty$ (with $K$ fixed), then the dynamic component gets pushed beyond the value of $K$, and only the static component contributes:
\begin{equation}
\label{eq:limit2}
\lim_{K\rightarrow\infty} \lim_{v\rightarrow \infty} \sum_{m=1}^{K} P(n^B \vert m^A) =
\lim_{K\rightarrow\infty} \sum_{m=1}^{K} O(n^B \vert m^A) =
\frac{1}{r} - \frac{\sin(2\pi n /r)}{2\pi n}.
\end{equation}
The physical interpretation should be clear.
For any fixed piston speed $v$, the sudden approximation breaks down if $m^A\pi/A \gtrsim v$;
for such initial states the evolving wavefunction catches up with the moving piston.
Therefore if we sum over {\it all} initial states at {\it fixed} $v$, then this sum necessarily includes states that violate the sudden approximation.
Conversely, the use of the sudden approximation in Eq.~\ref{eq:questionableLimit} is equivalent to imposing a cutoff $K$ on the sum over initial states: the effect of this cutoff is to exclude those states that give rise to the dynamic component, $s_n^R$.

The result appearing in Eq.~\ref{eq:limit2} is the same as that obtained for the process of sudden expansion into a vacuum, in which the length of the box increases {\it instantaneously} from $A$ to $B$.
This case, considered explicitly by TM (see Eq.~23 of Ref.~\cite{Teifel2010}) and for the classical piston by Sung~\cite{Sung2008}, highlights the importance of the ordering of limits for the validity of Eq.~\ref{eq:nwr}.
This issue is discussed in detail by Press\' e and Silbey~\cite{Presse2006}. See also Kurchan's lectures~\cite{Kurchan2009} for an alternative analysis of the sudden expansion process.

\begin{figure}[tbp]
   \begin{center}
      \subfigure[$\,\,v=10$]{
         \label{fig:PP_v=10}
         \includegraphics[trim = 0in 0in 0in 0in , scale=0.28,angle=270]{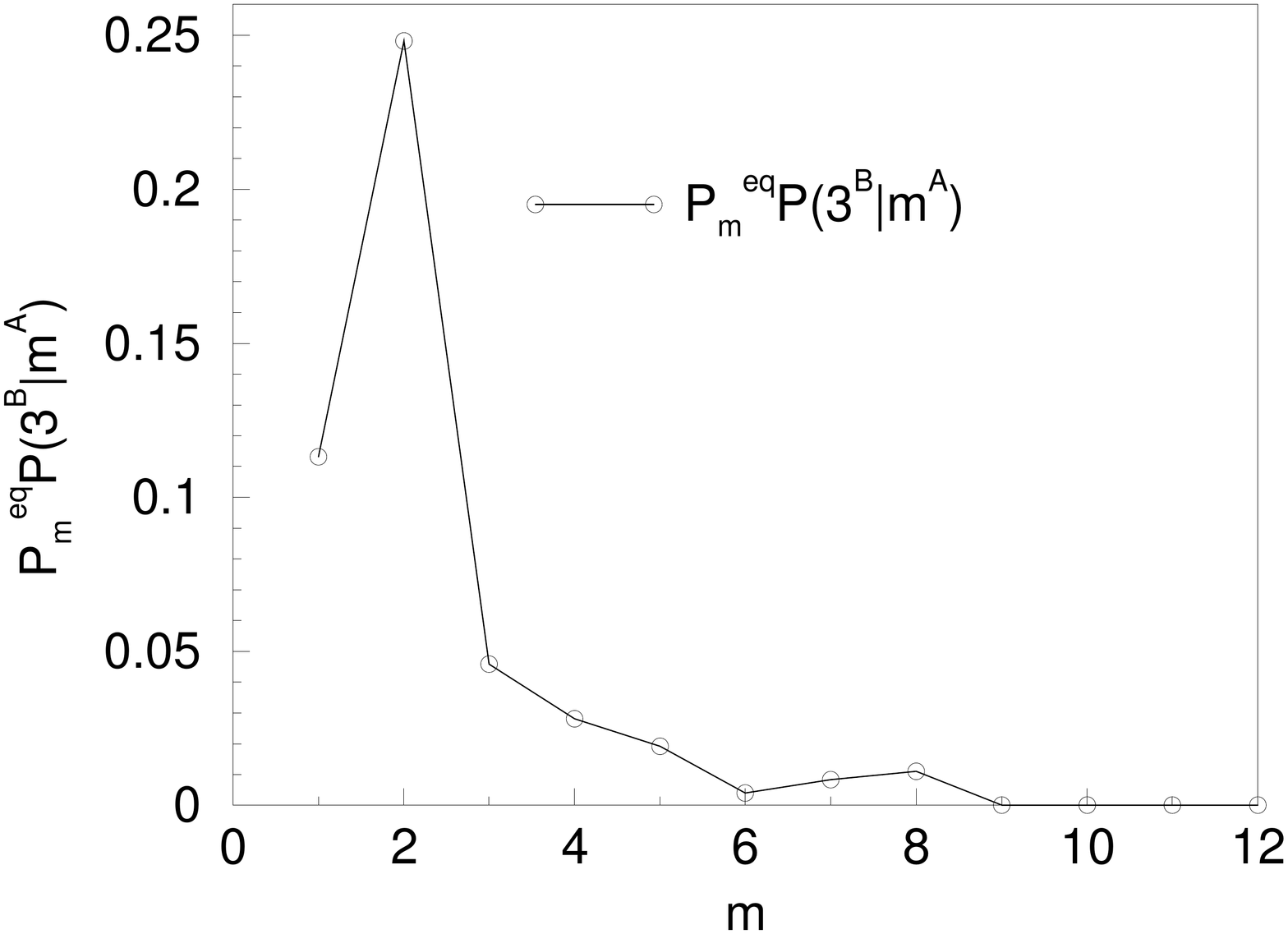}
      }
      \subfigure[$\,\,v=100$]{
         \label{fig:PP_v=100}
         \includegraphics[trim = 0in 0in 0in 0in , scale=0.28,angle=270]{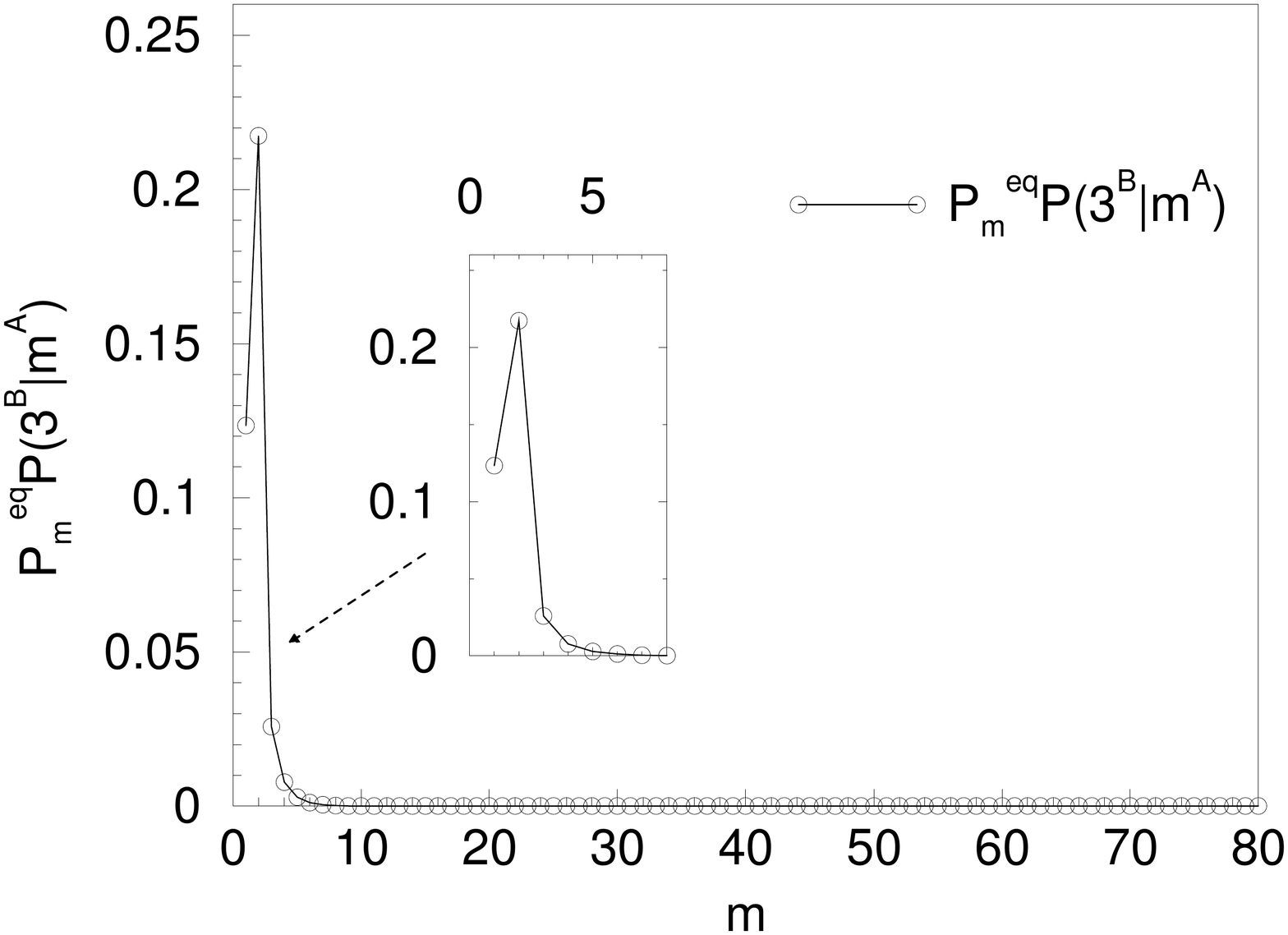} 
      } 
      \subfigure[$\,\,v=500$]{
         \label{fig:PP_v=500}
         \includegraphics[trim = 0in 0in 0in 0in , scale=0.28,angle=270]{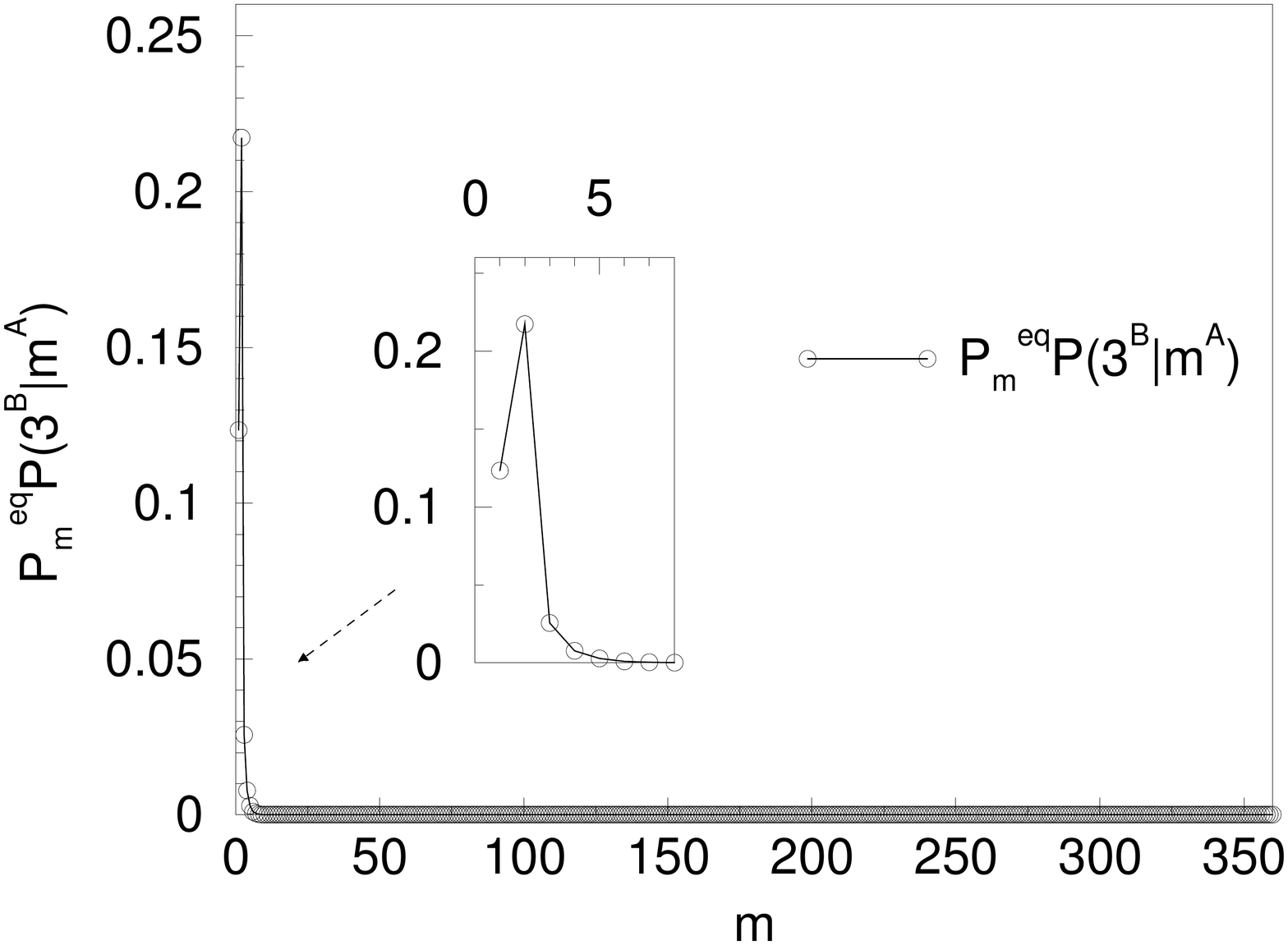} 
      } 
   \end{center}
   \caption{The probability to generate a realization from $\vert m^A\rangle$ to $\vert 3^B\rangle$ is plotted, for the same parameters as in Fig.~\ref{fig:peaks}, and taking $\beta = 0.01$.}
    \label{fig:PP}
\end{figure}

While our arguments establish that Eq.~\ref{eq:nwr} is valid for any piston expansion speed $v$,
they also imply that for large $v$, transitions $\langle n^B \vert \hat U \vert m^A \rangle$ from high-lying initial energy eigenstates make a large contribution to $s_n$ and ultimately to $\langle e^{-\beta W}\rangle$ (Eq.~\ref{eq:standardDerivation}).
When the energies of such high-lying states are much greater than $\beta^{-1}$, then the probability to sample these states from the initial canonical distribution, $P_m^{{\rm eq},A}\propto e^{-\beta E_m^A}$, becomes exceedingly small.
In this case, even though Eq.~\ref{eq:nwr} is valid, the number of realizations required to confirm its validity is prohibitively large.
Fig.~\ref{fig:PP} illustrates this point by displaying the product $P_m^{{\rm eq},A} \, P(n^B \vert m^A)$, that is the net probability to generate a realization with initial and final states $\vert m^A\rangle$ and $\vert n^B\rangle$, respectively, setting $\beta=0.01$ and $n=3$.
Comparing Figs.~\ref{fig:peaks} and \ref{fig:PP}, we see that although realizations that correspond to the dynamic component represent an important contribution to $s_n$, the probability to observe these realizations is vanishingly small.
This conclusion is mirrored in the classical version of this expanding piston~\cite{Lua2005,Jarzynski2006}, where a substantial contribution to $\langle e^{-\beta W}\rangle$ arises from single-collision events, in which the particle loses energy as it strikes the rapidly receding piston.
If $Mv^2 \gg \beta^{-1}$, then many realizations of the process are needed in order to stand a decent chance of sampling initial conditions in which the particle is moving sufficiently fast to collide with the piston.
By analogy with the classical calculations of Ref.~\cite{Lua2005,Jarzynski2006}, we expect that the number of realizations needed for the convergence of the exponential average in Eq.~\ref{eq:nwr} scales like $\exp(\beta M v^2)$, for large $v$.

We note in passing that in Figs.~\ref{fig:v=100} and \ref{fig:v=500} the right peak itself exhibits a double-peak structure.
This too has a semiclassical interpretation, which is easiest to explain in terms of the compression process.
At $t=0$ in the state $\vert 3^B\rangle$, the particle is moving with speed $\vert u\vert \approx 5$ (Eq.~\ref{eq:u=5}).
Its speed after a collision with the leftward-moving piston is greater if the particle was moving toward the piston just before the collision ($u\approx +5$) than if it was moving away from the piston ($u\approx -5$).
A back-of-the envelope calculation suggests that this difference splits the right peak into two sub-components separated by $\Delta m = 2 A \vert u\vert / \pi \approx 3$, in agreement with what we see in Figs.~\ref{fig:v=100} and \ref{fig:v=500}.

\begin{figure}[tbp]%
\begin{center}
      \subfigure[$\,\,v=0.1$]{%
         \label{fig:v=0.1}%
        \includegraphics[width=7.5cm, clip]{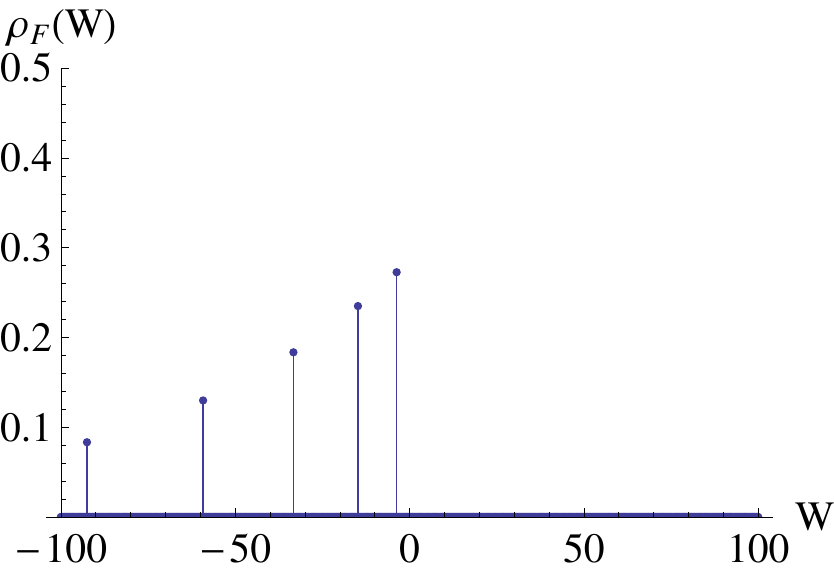}
      }%
      \subfigure[$\,\,v=1$]{%
         \label{fig:v=1}%
         \includegraphics[width=7.5cm, clip]{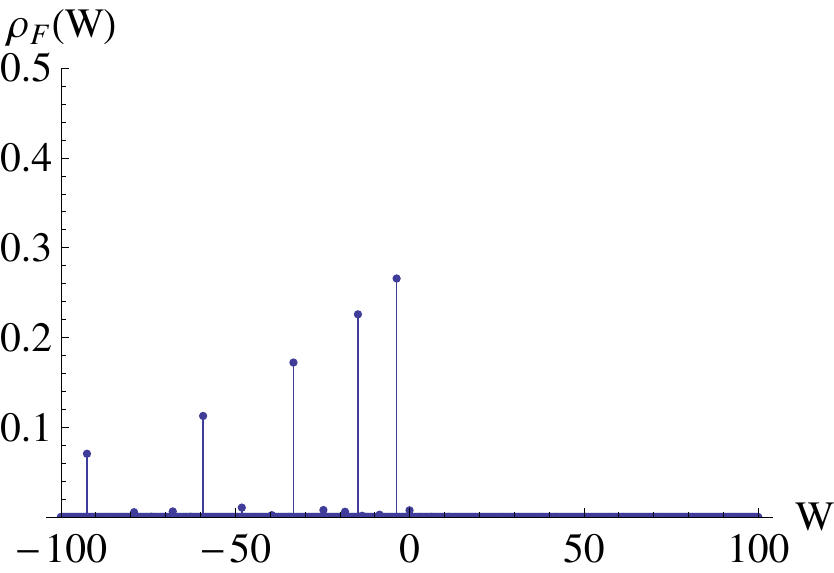}
      } \ \
      \subfigure[$\,\,v=2$]{%
         \label{fig:v=2}
         \includegraphics[width=7.5cm, clip]{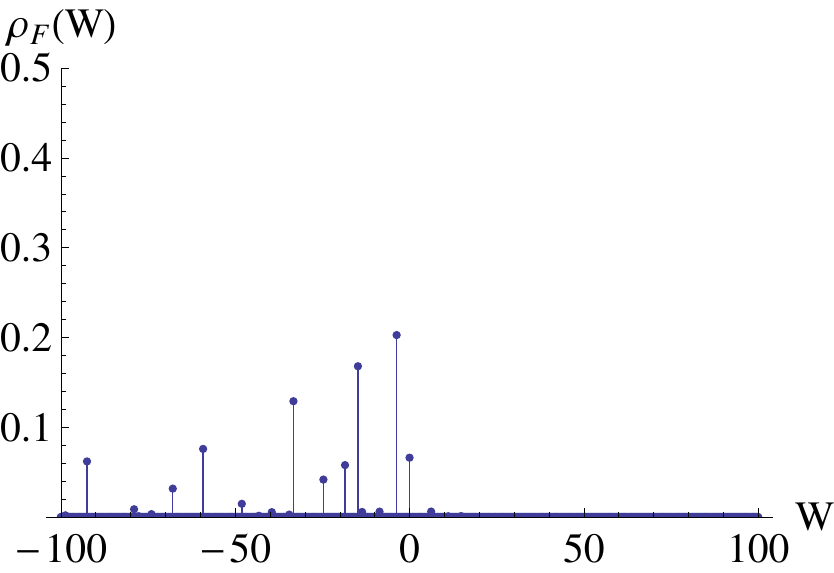}
      } 
        \subfigure[$\,\,v=4$]{
         \label{fig:v=4}
        \includegraphics[width=7.5cm, clip]{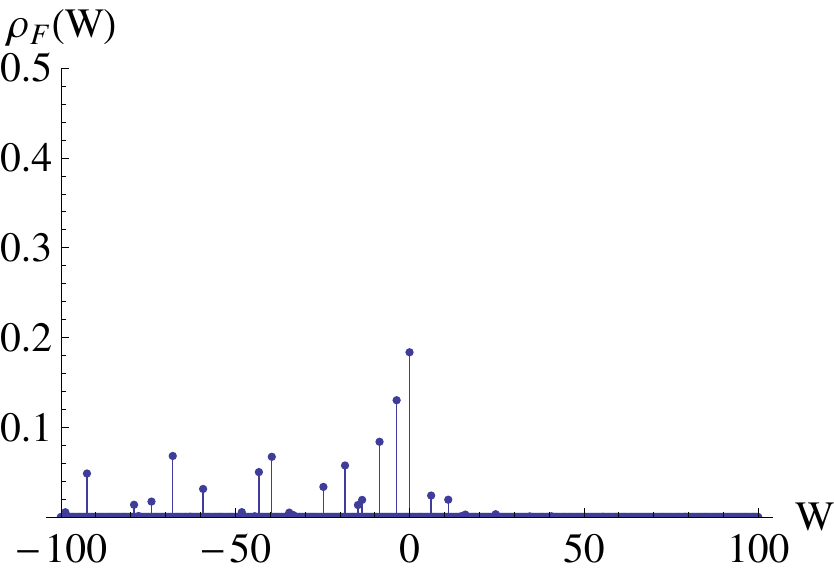}
      }
      \subfigure[$\,\,v=8$]{
         \label{fig:v=8}
         \includegraphics[width=7.5cm, clip]{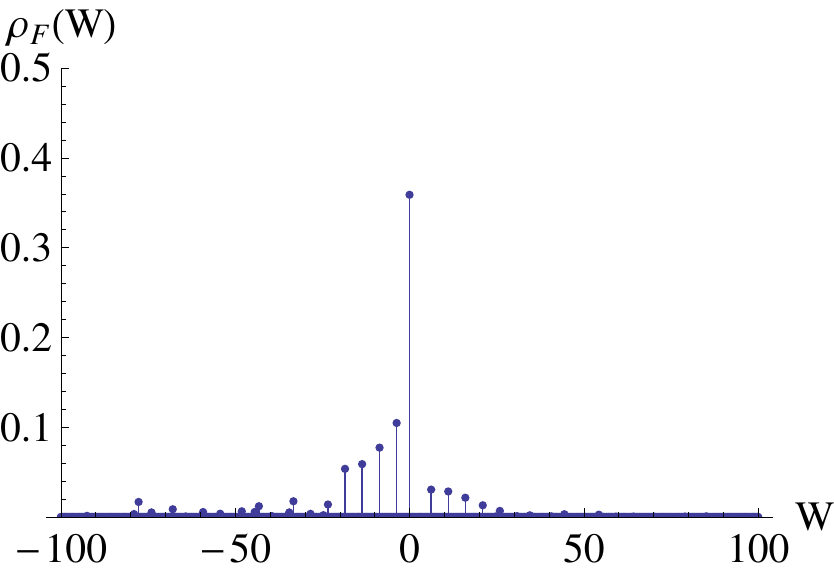}
      } 
      \subfigure[$\,\,v=1000$]{
         \label{fig:v=1000}
         \includegraphics[width=7.5cm, clip]{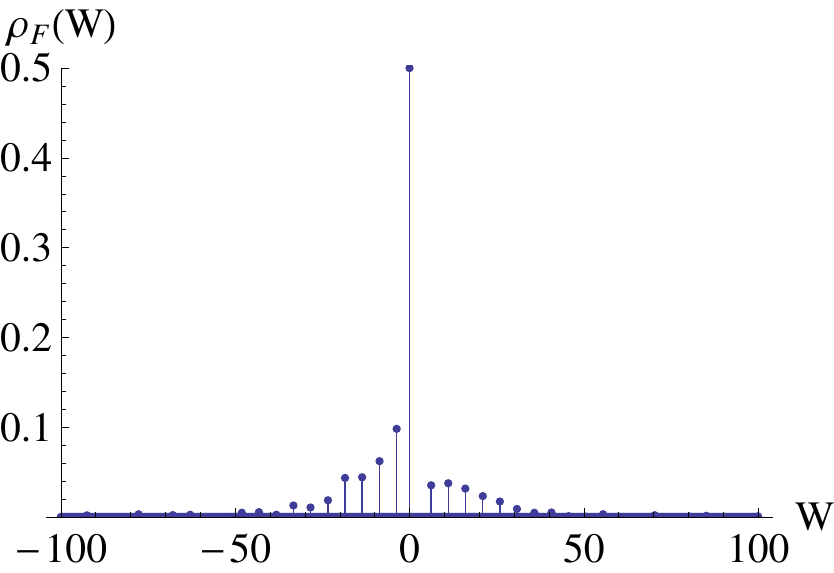}
      } 
      \end{center}
   \caption{Work distribution for the expanding quantum piston.
Here $A=1$,  $B=2$, $\beta=0.01$, and the piston speed ranges from $v=0.1$ to $v=1000$. The free energy difference is $\Delta
F=-\beta^{-1} \ln (B/A)\approx -30.10$.  The left tail of the distributions in the region $W<-100$ is not shown.}
    \label{fig:workdistribution}%
\end{figure}

Finally, since this model provides a useful pedagogical illustration of a quantum nonequilibrium process (see also Refs.~\cite{Quan2008,Deffner2008,Deffner2010,Talkner2008}), we briefly discuss the work distribution $\rho_F(W)$ for the expanding quantum piston (see Eq.~\ref{eq:derivecft}), plotted in Fig.~\ref{fig:workdistribution} for various piston speeds.
In the limit $v\rightarrow 0$, the quantum adiabatic theorem gives us $P(n^B\vert m^A) \rightarrow \delta_{mn}$.
Thus the work distribution in Fig.~\ref{fig:v=0.1} reflects the initial thermal energy distribution:
the largest peak corresponds to the situation in which the system begins and ends in the ground state,
the next largest corresponds to the first excited state, and so on.
In the opposite limit of large $v$, $\rho_F(W)$ approaches an asymptotic distribution, obtained by replacing $P(n^B\vert m^A)$ with its static component $O(n^B\vert m^A)$ in Eq.~\ref{eq:derivecft}.
(However, the dynamic component, which gets pushed off to infinity as discussed earlier, remains essential for the the validity of Eq.~\ref{eq:nwr}.)
There are two uniquely quantal features of the distributions shown in Fig.~\ref{fig:workdistribution}.
First, for $v\ge 2$ we can clearly see a nonzero probability to obtain a positive value of work.
This is forbidden in the classical case, as the particle loses energy each time it collides with the piston.
Second, for the classical expanding piston the probability to obtain $W=0$ approaches unity as $v\rightarrow\infty$,
whereas for the quantum piston with $A=1.0$ and $B=2.0$ this probability approaches 1/2, as illustrated by the peak at $W=0$ in Fig.~\ref{fig:v=1000}.
Finally, although it might not be obvious from Fig.~\ref{fig:workdistribution}, the average work performed in the limit $v\rightarrow\infty$ is zero for the quantum piston~\cite{Schlitt1970}, just as it is for the classical piston.

To conclude, we have used exact solutions of the time-dependent Schr\" odinger equation to study the validity of nonequilibrium work relations (Eqs.~\ref{eq:nwr}, \ref{eq:cft}) for the quantum piston, focusing on the limit of a rapidly expanding piston, $v\rightarrow\infty$.
Our investigation was motivated by Teifel and Mahler's study~\cite{Teifel2010}, which highlighted the subtleties that arise when the system's Hilbert space changes due to the motion of hard boundaries.
As in the classical case, we found that both Eqs.~\ref{eq:nwr} and \ref{eq:cft} remain valid for any finite piston speed, but the convergence of $\left\langle e^{-\beta W}\right\rangle$ to $e^{-\beta \Delta F}$ requires a sum over {\it all} possible realizations.
In particular, when $v \gg \beta^{-1/2}$ important contributions arise from those rare realizations in which the particle begins with a sufficiently high energy to collide with the piston.
These realizations show up as the dynamic component (the right peak) in Fig.~\ref{fig:peaks}.
Although we have considered only the one-dimensional quantum piston, we speculate that similar conclusions will apply to more complicated quantum systems involving moving hard boundaries, for which exact solutions of the Schr\" odinger equation are unavailable.

\acknowledgments
We gratefully acknowledge support from the National Science Foundation (USA) under grant DMR-0906601.
HTQ thanks Prof.\ Jaeyoung Sung for stimulating discussions and Andy Ballard for help with computational matters.

\appendix

\section{}

Eq.~\ref{eq:exactSoln} gives the following expression for the transition probability from $\vert m^A\rangle$ to $\vert n^B\rangle$ during the expansion process:
\begin{equation}
\begin{split}
&P(n^B \vert m^A) =\left \vert \sum_{l=1}^{\infty}\frac{2}{A} \int_{0}^{A} e^{-i v x^2/2 A}  \sin \left(\frac{l \pi x}{A}\right)  \sin \left(\frac{m \pi x}{A}\right)   {\rm d}x \right.\\
&\times \left.  
\exp \left[ -i \frac{\pi^2 l^2 (B-A)}{2ABv} \right] \,
\frac{2}{B} \int_{0}^{B}  e^{i v x^2/2 B}  \sin \left( \frac{n \pi
x}{B} \right) \sin \left( \frac{l \pi x}{B} \right) {\rm d}x \right \vert^{2}.
\end{split}
\label{expansion}
\end{equation}
For the contraction process, the transition probability from $\vert n^B\rangle$ to $\vert m^A\rangle$ is obtained from this result by making the replacements $m \leftrightarrow n$,
$A \leftrightarrow B$, and $v \rightarrow -v$:
\begin{equation}
\begin{split}
&\bar{P}(m^A \vert n^B) =\left \vert \sum_{l=1}^{\infty}\frac{2}{B} \int_{0}^{B} e^{i v x^2/2 B}  \sin \left(\frac{l \pi x}{B}\right)  \sin \left(\frac{n \pi x}{B}\right)   {\rm d}x \right.\\
&\times \left.  
\exp \left[ i \frac{\pi^2 l^2 (A-B)}{2BAv} \right] \,
\frac{2}{A} \int_{0}^{A}  e^{-i v x^2/2 A}  \sin \left( \frac{m \pi
x}{A} \right) \sin \left( \frac{l \pi x}{A} \right) {\rm d}x \right \vert^{2}.
\end{split}
\label{contraction}
\end{equation}
Comparing these expressions, it is straightforward to verify that they are equal:
\begin{equation}
P(n^B \vert m^A) = \bar{P}(m^A \vert n^B)
\label{a3}
\end{equation}

\bibliography{CJ_references}

\end{document}